\documentclass[11pt, notitlepage]{article}
\pdfoutput=1

\usepackage{float}
\usepackage{subfig}
\usepackage{graphicx}
\usepackage[T1]{fontenc}
\usepackage[utf8]{inputenc}
\usepackage{authblk}
\usepackage{amsmath,bm}
\usepackage{amssymb}
\usepackage{amsmath}
\usepackage{enumitem}
\usepackage{mathtools}
\usepackage[left=2cm,right=2cm,top=2cm,bottom=2cm]{geometry}

\renewcommand\footnotemark{}
\begin{document}

\title{\bf{Is Asymptotically Weyl-Invariant Gravity Viable?}}
\author{Daniel Coumbe}
\thanks{E-mail: dc@norreg.dk}
\affil{\small{\emph{The Niels Bohr Institute, Copenhagen University}\\ \emph{Blegdamsvej 17, DK-2100 Copenhagen Ø, Denmark}}}
\date{}
\maketitle


\begin{abstract}

  We explore the cosmological viability of a theory of gravity defined by the Lagrangian $f(\mathcal{R})=\mathcal{R}^{n\left(\mathcal{R}\right)}$ in the Palatini formalism, where $n\left(\mathcal{R}\right)$ is a dimensionless function of the Palatini scalar curvature $\mathcal{R}$ that interpolates between general relativity when $n\left(\mathcal{R}\right)=1$ and a locally scale-invariant and superficially renormalizable theory when $n\left(\mathcal{R}\right)=2$. We refer to this model as asymptotically Weyl-invariant gravity (AWIG).

  We analyse perhaps the simplest possible implementation of AWIG. A phase space analysis yields three fixed points with effective equation of states corresponding to de Sitter, radiation and matter-dominated phases. An analysis of the deceleration parameter suggests our model is consistent with an early and late period of accelerated cosmic expansion, with an intermediate period of decelerated expansion. We show that the model contains no obvious curvature singularities. Therefore, AWIG appears to be cosmologically viable, at least for the simple implementation explored.
  

\vspace{0.25cm}

\end{abstract}


\begin{section}{Introduction}

  The general theory of relativity is currently our best description of gravity. One reason for this is its explanatory power: assuming little more than a single symmetry principle general relativity can explain a truly astonishing range of experimental phenomena~\cite{Will:2014kxa}.

  However, it is at best incomplete. It is often said that general relativity breaks down at high energies \emph{or} small distances. Yet, it is more accurate to say high energies \emph{and} small distances. This is an important distinction since it highlights the regime in which we must modify general relativity, namely for large energy densities, or equivalently for large spacetime curvatures. For example, general relativity predicts its own breakdown at curvature singularities, where scalar measures of curvature grow without bound. Furthermore, general relativity is known to become fundamentally incompatible with quantum field theory at high curvature scales, a failure known as its non-renormalizability~\cite{'tHooft:1974bx,Goroff:1985th}. Theoretical arguments alone are enough to tell us that general relativity must be modified at high curvature scales.
  
  Experimental data also indicates that general relativity must either be augmented or replaced altogether if it is to agree with observation~\cite{Akrami:2018odb}. For example, general relativity by itself is unable to explain the early phase of accelerated cosmic expansion, as evidenced by myriad high-precision measurements~\cite{Akrami:2018odb}, and must be supplemented with unobserved exotic energy sources and scalar fields~\cite{Carroll:2000fy,Copeland:2006wr}. Although this top-down approach, as exemplified by the $\Lambda$CDM model, is currently our best description of observed cosmological dynamics~\cite{Akrami:2018odb}, its \emph{ad hoc} construction has driven attempts to replace general relativity from the bottom-up.

  Finding a viable replacement of general relativity is challenging. Such a theory must at the very least be (\romannumeral 1) equivalent to general relativity in the low-curvature limit, (\romannumeral 2) renormalizable in the high-curvature limit, (\romannumeral 3) unitary, (\romannumeral 4) stable, (\romannumeral 5) contain no curvature singularities, (\romannumeral 6) consistent with observation.

  One attempt is that of higher-order gravity, in which the Lagrangian includes terms quadratic in the curvature tensor. Although this approach is perturbatively renormalizable, and hence satisfies criterion (\romannumeral 2), such higher-order theories are not typically unitary or stable, thus failing to satisfy criteria (\romannumeral 3) and (\romannumeral 4). The only higher-order theories that are unitary and stable are so-called $f(R)$ theories, in which the Lagrangian is an arbitrary function of the Ricci scalar only~\cite{Sotiriou:2008rp}.

  There are three types of $f(R)$ theory: metric, Palatini and metric-affine variations~\cite{Sotiriou:2008rp}. Metric $f(R)$ gravity assumes that the affine connection uniquely depends on the metric via the Levi-Civita connection, as in standard general relativity. The Palatini formalism generalises the metric formalism by relaxing the assumption that the connection must depend on the metric. The metric-affine formalism is the most general approach since it even drops the implicit assumption that the matter action is independent of the connection. 

  Particular metric $f(R)$ models have been shown to conflict with solar system tests~\cite{Chiba:2003ir}, give an incorrect Newtonian limit~\cite{Sotiriou:2005hu}, contradict observed cosmological dynamics~\cite{Amendola:2006kh,Amendola:2006we}, be unable to satisfy big bang nucleosynthesis constraints~\cite{Brookfield:2006mq} and contain fatal Ricci scalar instabilities~\cite{Dolgov:2003px}. Thus, metric $f(R)$ theories do not typically satisfy criteria (\romannumeral 1), (\romannumeral 4) or (\romannumeral 6). As for the metric-affine formalism, it is not even a metric theory in the usual sense, meaning diffeomorphism invariance is likely broken~\cite{Sotiriou:2008rp}. Thus, metric-affine theories do not even seem to satisfy criterion (\romannumeral 1). However, it has been shown that the Palatini variation is immune to any such Ricci scalar instability~\cite{Meng:2003sxc}. Palatini formulations also appear to pass solar system tests and reproduce the correct Newtonian limit~\cite{Sotiriou:2005xe}. Remarkably, a Palatini action that is linear in the scalar curvature is identical to regular general relativity~\cite{Sotiriou:2008rp}. However, this equivalence does not hold for higher-order theories~\cite{BeltranJimenez:2017vop,Sotiriou:2008rp}. In particular, a Palatini action that is purely quadratic in the scalar curvature is identical to normal general relativity plus a non-zero cosmological constant~\cite{Edery:2019txq}. 

  In Ref.~\cite{Coumbe:2019fht} we proposed the theory of asymptotically Weyl-invariant gravity (AWIG) within the Palatini formalism (see Refs.~\cite{Coumbe:2015zqa,Coumbe:2015aev,Coumbe:2018myj} for the background to this proposal). By construction AWIG satisfies criteria (\romannumeral 1)-(\romannumeral 4).\interfootnotelinepenalty=10000 \footnote{\scriptsize In the low-curvature limit AWIG yields $f \left(\mathcal{R}\right)=\mathcal{R}$, which is identical to general relativity~\cite{Sotiriou:2008rp}. AWIG is at least superficially renormalizable because the coupling constant of the theory becomes dimensionless in the high curvature limit, as shown in section~\ref{model}. AWIG is likely to be unitary because states of negative norm (ghosts) that cause unitarity violations do not appear in $f(R)$ theories~\cite{Sotiriou:2008rp,Stelle:1977ry}. AWIG appears stable since Ostragadsky’s instability is evaded by any $f(R)$ theory~\cite{Woodard:2006nt}, and the Dolgov-Kawasaki instability can not occur in Palatini $f\left(\mathcal{R}\right)$ gravity~\cite{Sotiriou:2008rp} (see Ref.~\cite{Coumbe:2019fht} for more details on the construction of AWIG).} The present work aims to test whether this theory also satisfies criteria (\romannumeral 5) and (\romannumeral 6), and hence to determine if it may be a viable replacement of general relativity.

 In addition to satisfying criteria (\romannumeral 1)-(\romannumeral 4), a major motivation for developing AWIG was finding a theory with the symmetry of local scale invariance. The need for local scale invariance can be seen by recognising that all length measurements are local comparisons. For example, to measure the length of a rod requires bringing it together with some standard unit of length, say a metre stick, at the same point in space and time. In this way the local comparison yields a dimensionless ratio, for example, the rod might be longer than the metre stick by a factor of two. Repeating this comparison at a different spacetime point must yield the same result, even if the metric at this new point were rescaled by an arbitrary factor $\Omega^{2}(x)$. This is because both the rod and metre stick would be equally rescaled, yielding the same dimensionless ratio. Such a direct comparison cannot be made for two rods with a non-zero space-like or time-like separation~\cite{Weyl:1918ib,PhysRev.125.2163}. Therefore, it has been argued that the laws of nature must be formulated in such a way as to be invariant under local rescalings of the metric tensor $g_{\mu\nu} \rightarrow \Omega^{2}(x) g_{\mu\nu}$, or equivalently under a local change of units. Moreover, since scale-invariant theories of gravity are gauge theories~\cite{Wesson,Lasenby:2015dba}, unification with the other three fundamental interactions, which have all been successfully formulated as local gauge theories, becomes tractable. The theory analysed in this work is invariant with respect to local changes of scale in the high-curvature limit.   

  It is important to establish the standard against which we will judge whether the presented theory is viable. Criterion (\romannumeral 5) will be deemed to be satisfied if at least two different curvature invariants can be shown to be divergence-free. To satisfy criterion (\romannumeral 6) we make the maximal demand that the theory reproduces all four observed phases of cosmological evolution in the correct order~\cite{Fay:2007gg}, namely an early period of accelerated expansion, followed by radiation and matter-dominated phases, and finally a late period of accelerated expansion~\cite{Spergel:2006hy}.

  This paper is organised as follows. In section~\ref{model} we define the model of AWIG, including a detailed exploration of the dimensionless exponent $n\left(\mathcal{R}\right)$. In section~\ref{method} we detail the methodology that will be used to test the viability of our model. Results are presented in section~\ref{results} followed by a concluding discussion in section~\ref{discussion}.   

\end{section}


\begin{section}{Model}\label{model}

The class of theories to which our model belongs is defined by the action
  
\begin{equation}\label{a0}
\mathcal{S}=\frac{1}{2\kappa} \int f \left(\mathcal{R}\right) \sqrt{-g}d^{4}x,
\end{equation}

\noindent where $\kappa\equiv 8\pi G$ and $G$ is the gravitational coupling. $f \left(\mathcal{R}\right)$ is an arbitrary function of the Palatini scalar curvature $\mathcal{R}$ and $g$ is the determinant of the metric tensor. Varying Eq.~(\ref{a0}) with respect to the metric and taking the trace gives the field equations~\cite{Sotiriou:2008rp}

\begin{equation}\label{traceFE}
f'(\mathcal{R}) \mathcal{R}- 2f(\mathcal{R})=\kappa T.
\end{equation}

AWIG is defined by the specific case~\cite{Coumbe:2019fht}

\begin{equation}\label{f1}
f\left(\mathcal{R}\right)=\mathcal{R}^{n\left(\mathcal{R}\right)},
\end{equation}

\noindent where $n\left(\mathcal{R}\right)$ is a dimensionless function of $\mathcal{R}$ that interpolates between general relativity when $n\left(\mathcal{R}\right)=1$ and a locally scale-invariant and superficially renormalizable theory of gravity when $n\left(\mathcal{R}\right)=2$. By defining $n\left(\mathcal{R}\right)$ in this way the Lagrangian density $f\left(\mathcal{R}\right)$ is purely a function of scalar curvature, and hence is guaranteed to be invariant under arbitrary differential coordinate transformations. In $4$-dimensional spacetime $\mathcal{R}^{n\left(\mathcal{R}\right)}$ has canonical mass dimension $2n\left(\mathcal{R}\right)$. Since $\sqrt{-g}$ has mass dimension $-4$, $\kappa$ must have a mass dimension of $2n\left(\mathcal{R}\right)-4$ if Eq.~(\ref{a0}) is to be dimensionless, which it must be since we are working in units of $\hbar=c=1$. Thus, in the limit $n\left(\mathcal{R}\right) \to 2$ the gravitational coupling becomes dimensionless, as demanded by scale-invariance. Superficially renormalizable field theories are those with dimensionless coupling constants~\cite{Kawai:1995ju}. 

To complete the definition of this model we must specify the function $n\left(\mathcal{R}\right)$. We begin by taking the first derivative of $f\left(\mathcal{R}\right)$ with respect to $\mathcal{R}$, denoted by $f'\left(\mathcal{R}\right)$, finding 

\begin{equation}\label{fp1}
f'\left(\mathcal{R}\right) = \mathcal{R}^{n\left(\mathcal{R}\right)-1}\left(n\left(\mathcal{R}\right) + \mathcal{R}\rm{log}\left(\mathcal{R}\right)n'\left(\mathcal{R}\right)\right).
\end{equation}  

\noindent Substituting Eqs.(\ref{f1}) and (\ref{fp1}) into Eq.~(\ref{traceFE}) and rearranging yields

\begin{equation}\label{fp2}
  n'\left(\mathcal{R}\right)=\frac{\kappa T+\mathcal{R}^{n\left(\mathcal{R}\right)}\left(2-n\left(\mathcal{R}\right)\right)}{\mathcal{R}^{n\left(\mathcal{R}\right)+1}\log{\left(\mathcal{R}\right)}}.
\end{equation} 

We now use the fact that the symmetry of local scale invariance is signalled by the vanishing of the traced energy tensor~\cite{Mukhanov:2007zz}. Thus, as $n\left(\mathcal{R}\right)\to 2$ we must have $T \to 0$. Applying the limits $n\left(\mathcal{R}\right)\to 2$ and $T \to 0$ to Eq.(\ref{fp2}) yields $n'\left(\mathcal{R}\right)=0$. Similarly, as $n\left(\mathcal{R}\right)\to 1$ we must have $\kappa T\to -\mathcal{R}$, and so Eq.(\ref{fp2}) again yields $n'\left(\mathcal{R}\right)=0$.\interfootnotelinepenalty=10000 \footnote{\scriptsize If $\mathcal{R}=1$ when $n\left(\mathcal{R}\right)=2$ and $\kappa T=0$ then $n'\left(\mathcal{R}\right)$ is undefined, since the numerator and denominator of Eq.(\ref{fp2}) both equal zero. Likewise, if $\mathcal{R}=0$ when $n\left(\mathcal{R}\right)=1$ and $\kappa T=-\mathcal{R}$ then $n'\left(\mathcal{R}\right)$ is undefined. However, in the limiting cases $\mathcal{R}\to 0$ and $\mathcal{R}\to 1$ we have $n'\left(\mathcal{R}\right)=0$.} Therefore, the function we seek must satisfy the condition $n'\left(\mathcal{R}\right)=0$ as $n\left(\mathcal{R}\right)\to 1$ and $n\left(\mathcal{R}\right)\to 2$.

Experiment also supports a near-constant exponent $n\left(\mathcal{R}\right)$ at lower curvature scales. This is because general relativity agrees with experiment over a wide range of energy or curvature scales~\cite{Clifton:2005aj,Will:2014kxa}, indicating that $n\left(\mathcal{R}\right)$ has at most a very weak dependence on $\mathcal{R}$ within the range of current experimental sensitivity. Similarly, the fact that in the high-curvature limit the theory becomes locally scale-invariant implies a constant $n\left(\mathcal{R}\right)$, since in this limit there can be no scale with respect to which $n\left(\mathcal{R}\right)$ can vary.            

We now proceed by assuming $n\left(\mathcal{R}\right)$ admits a series expansion in $\mathcal{R}$ of the form

\begin{equation}\label{a2}
n\left(\mathcal{R}_{*}\right)= \sum_{m=0}^{\infty}c_{m}\mathcal{R}_{*}^{m},
\end{equation}

\noindent where $c_{m}$ are dimensionless constants and $\mathcal{R_{*}}$ is defined by the dimensionless ratio $\mathcal{R_{*}}\equiv \mathcal{R}/ \mathcal{R}_{0}$, with $\mathcal{R}_{0}$ a finite constant of mass dimension two that represents the maximum value $R$ can take. In this way, $n\left(\mathcal{R}_{*}\right)$ is a purely dimensionless function of the Palatini scalar curvature $\mathcal{R}$. Truncating to a third-order function we have\interfootnotelinepenalty=10000 \footnote{\scriptsize It can be shown that first and second-order functions cannot produce the desired features~\cite{ebert2003texturing}.}

\begin{equation}\label{a3}
n\left(\mathcal{R}_{*}\right)= c_{0} + c_{1}\mathcal{R}_{*} + c_{2}\mathcal{R}_{*}^{2} + c_{3}\mathcal{R}_{*}^{3}.
\end{equation}

\noindent Since the low-curvature limit corresponds to $\mathcal{R}_{*}\to 0$, the constraint $n\left(\mathcal{R}_{*}\to 0\right)=1$ immediately yields $c_{0}=1$. Similarly, since the high-curvature limit corresponds to $\mathcal{R}_{*}\to 1$, the constraint $n\left(\mathcal{R}_{*}\to 1\right)=2$ gives $1+c_{1} + c_{2} + c_{3}=2$, or equivalently $c_{1} + c_{2} + c_{3}=1$. 

The first derivative of $n\left(\mathcal{R}_{*}\right)$ with respect to $\mathcal{R}$ is 

\begin{equation}\label{a4}
n'\left(\mathcal{R}_{*}\right)= \frac{c_{1}}{\mathcal{R}_{0}} + 2\frac{c_{2}}{\mathcal{R}_{0}^{2}}\mathcal{R} + 3\frac{c_{3}}{\mathcal{R}_{0}^{3}}\mathcal{R}^{2}=\frac{c_{1}}{\mathcal{R}}\mathcal{R}_{*}+2\frac{c_{2}}{\mathcal{R}}\mathcal{R}_{*}^{2}+3\frac{c_{3}}{\mathcal{R}}\mathcal{R}_{*}^{3}.
\end{equation} 

\noindent Since $\mathcal{R}_{*}\equiv \mathcal{R}/\mathcal{R}_{0} \to 0$ in the low-curvature limit, Eq.~(\ref{a4}) gives $n'\left(\mathcal{R}_{*}\right)=c_{1}/\mathcal{R}_{0}=0$, which implies $c_{1}=0$ since $\mathcal{R}_{0}$ is assumed to be finite. The high-curvature limit corresponds to $\mathcal{R}_{*}\equiv \mathcal{R}/\mathcal{R}_{0} \to 1$, and so Eq.~(\ref{a4}) gives $n'\left(\mathcal{R}_{*}\right)=c_{1}/\mathcal{R}_{0}+2c_{2}/\mathcal{R}_{0}+3c_{3}/\mathcal{R}_{0}=0$, which implies $2c_{2}+3c_{3}=0$ since $c_{1}=0$. The polynomial coefficients $c_{2}$ and $c_{3}$ can now be determined by solving the system of equations $2c_{2}+3c_{3}=0$ and $c_{2}+c_{3}=1$, with the result $c_{2}=3,c_{3}=-2$. Therefore, 

\begin{equation}\label{a5}
n\left(\mathcal{R}_{*}\right)=1+3\mathcal{R}_{*}^{2}-2\mathcal{R}_{*}^{3}.
\end{equation} 

Eq.~(\ref{a5}) is the lowest-order polynomial to satisfy our criteria, but there are potentially an infinite number of higher-order polynomial functions. Let $n_{i}\left(\mathcal{R}_{*}\right)$ label this set of polynomial functions, where the order of the polynomial is given by $2i+1$. One can then generalise Eq.~(\ref{a5}) to any higher-order using~\cite{ebert2003texturing}

\begin{equation}\label{GenSmooth}
n_{i}\left(\mathcal{R}_{*}\right)=1+\mathcal{R}_{*}^{i+1} \sum_{j=0}^{i}  {{i+j}\choose{j}} {{2i+1}\choose{i-j}} \left(-\mathcal{R}_{*}\right)^{j},\qquad i \in \mathbb{N}.
\end{equation}

\noindent The first thirteen functions generated by Eq.~(\ref{GenSmooth}) are shown in Fig.~\ref{ni} (left). The Lagrangian density in this case is then

\begin{equation}\label{Lagrangian}
f_{i}\left(\mathcal{R}\right)=\mathcal{R}^{n_{i}\left(\mathcal{R}_{*}\right)}=\left(\mathcal{R}_{0}\mathcal{R}_{*}\right)^{n_{i}\left(\mathcal{R}_{*}\right)}.
\end{equation}


\noindent For simplicity, we choose $\mathcal{R}_{0}$ to have the value of one when expressed in some particular unit of mass dimension two. For example, one possibility is $\mathcal{R}_{0}=1m_{P}$, where $m_{P}$ is the Planck mass. The term $\mathcal{R}_{0}$ then only acts to set the dimensionality of $f_{i}\left(\mathcal{R}\right)$. The first thirteen functions $f_{i}\left(\mathcal{R}\right)$ generated by applying Eq.~(\ref{GenSmooth}) to Eq.~(\ref{Lagrangian}) are shown in Fig.~\ref{ni} (middle), where we set $\mathcal{R}_{0}=1$ in some appropriate unit. Differentiating Eq.~(\ref{Lagrangian}) with respect to $\mathcal{R}$ gives the set of first derivative functions $f'_{i}\left(\mathcal{R}\right)$, with the first 13 shown in Fig.~\ref{ni} (right).

\begin{figure}[H]
  \centering
\scalebox{.45}{\includegraphics{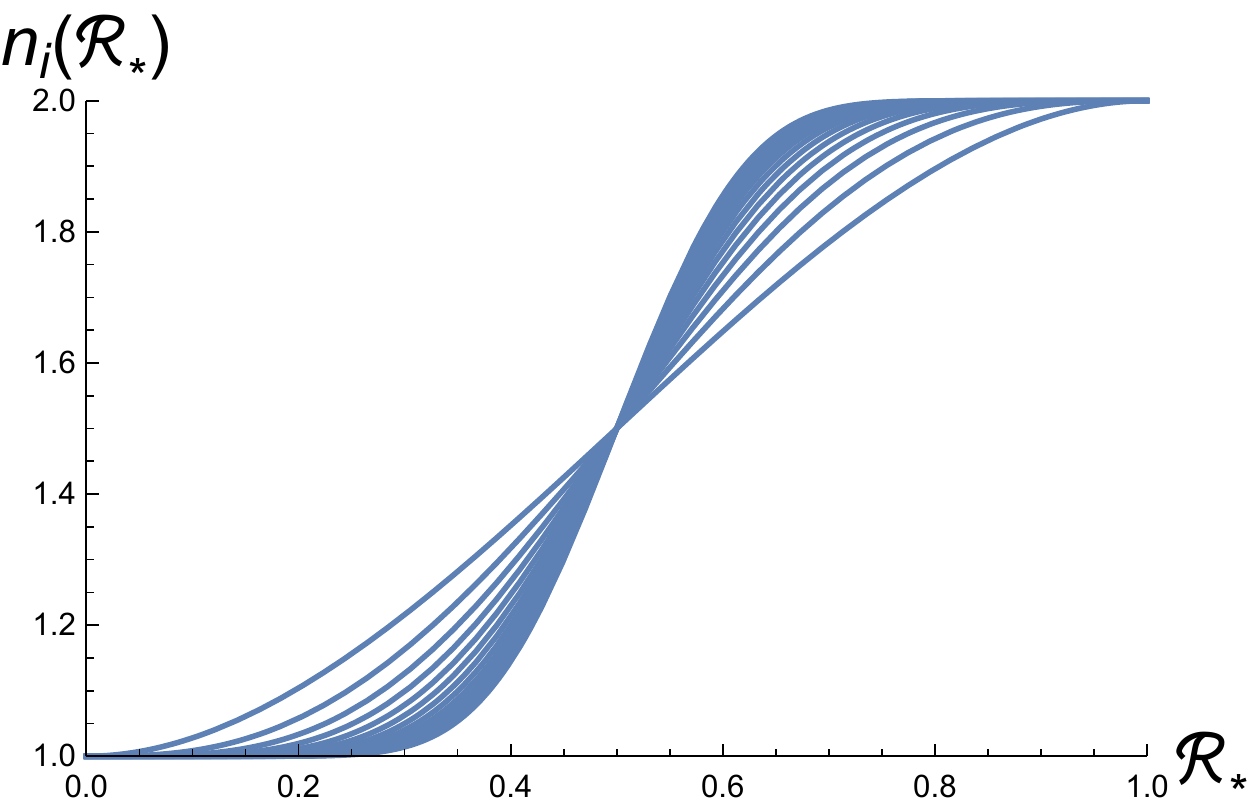}}
\scalebox{.45}{\includegraphics{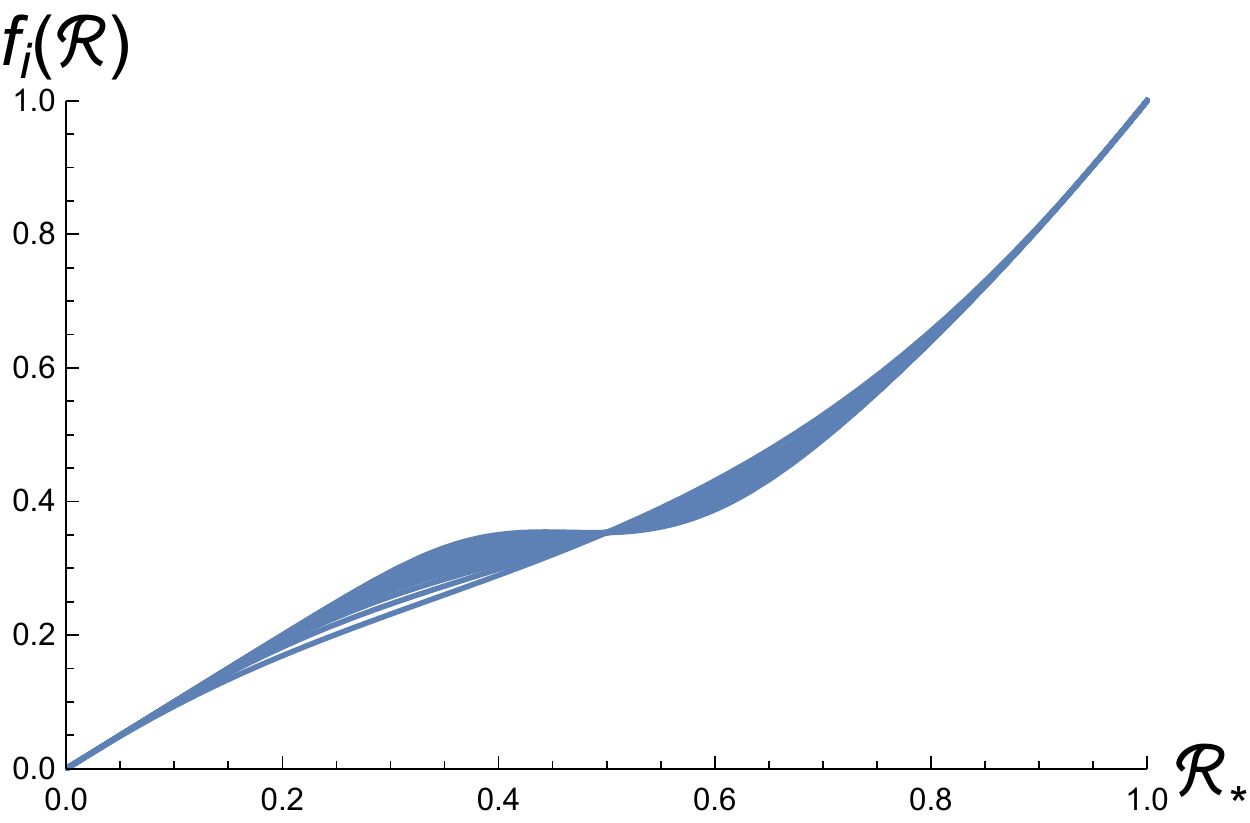}}
\scalebox{.45}{\includegraphics{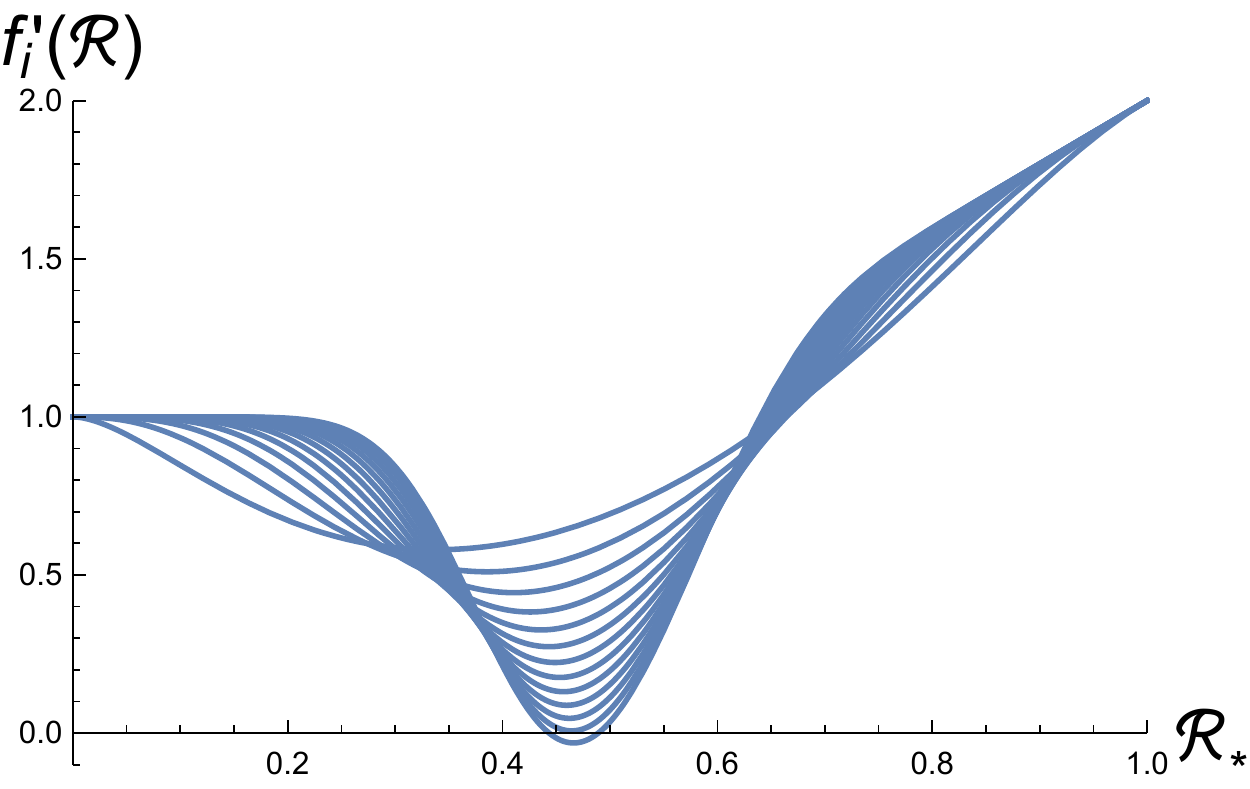}}
\caption{The first 13 exponents $n_{i}\left(\mathcal{R}_{*}\right)$ (left), Lagrangian densities $f_{i}\left(\mathcal{R}\right)$ (middle), and first derivative functions $f'_{i}\left(\mathcal{R}\right)$ (right) generated by Eq.~(\ref{GenSmooth}) as a function of $\mathcal{R}_{*}$.} 
\label{ni}
\end{figure}

An important feature of Fig.~\ref{ni} (right) is that the thirteenth function $f'_{13}\left(\mathcal{R}\right)$ becomes negative for certain values of $\mathcal{R}_{*}$. A well-defined conformal transformation of the metric tensor $\tilde{g}_{\mu\nu}=f'\left(\mathcal{R}\right)g_{\mu\nu}$ requires that $f'\left(\mathcal{R}\right)>0$ for all $\mathcal{R}$. This condition is only satisfied if $i\leq 12$. Thus, we can exclude Lagrangian densities $f_{i}\left(\mathcal{R}\right)$ with $i\geq 13$. In this work, we shall focus on the simplest permitted Lagrangian density

\begin{equation}\label{LagrangianSimple}
f_{1}\left(\mathcal{R}\right)=\mathcal{R}^{1+3\mathcal{R}_{*}^{2}-2\mathcal{R}_{*}^{3}}.
\end{equation}

\end{section}


\begin{section}{Method}\label{method}

  In this section we detail the method used to test the cosmological viablility of the model defined by Eq.~(\ref{LagrangianSimple}). The methodology presented in this section follows the work of Refs.~\cite{Fay:2007gg,Coumbe:2019fht}.

  Since cosmological observations by the Planck satellite show that our universe is consistent with being spatially flat at late times~\cite{Akrami:2018odb}, we begin by assuming a flat Friedmann-Lemaıtre-Robertson-Walker (FLRW) metric

\begin{equation}\label{m1}
ds^{2}=-dt^{2}+a^{2}(t)\left(dx^{2}+dy^{2}+dz^{2}\right),
\end{equation}

\noindent where $a(t)$ is the scale factor of the universe, a function of cosmological time $t$. The evolution of a spatially homogenous and isotropic universe filled with a cosmological fluid composed of pressureless dust and radiation can be described by the modified Friedmann equation~\cite{Sotiriou:2008rp,Cao:2017jsi}

\begin{equation}\label{m2}
\left(H+\frac{\dot f'\left(\mathcal{R}\right)}{2f'\left(\mathcal{R}\right)}\right)^{2}=\frac{\kappa\left(\rho_{m} + 2\rho_{r}\right)+f\left(\mathcal{R}\right)}{6f'\left(\mathcal{R}\right)},
\end{equation}

\noindent where the dot notation signifies a time derivative and $H\equiv \dot a/ a$ is the Hubble parameter. $\rho_{m}$ and $\rho_{r}$ are the energy density of matter and radiation, respectively, which satisfy the conservation conditions

\begin{equation}\label{cons}
\dot \rho_{m} + 3H\rho_{m}=0,\qquad \dot \rho_{r} + 4H\rho_{r}=0.
\end{equation}

Since the trace of the energy-momentum tensor for radiation is zero, we simply have $T=-\rho_{m}$~\cite{Fay:2007gg}. By using Eq.~(\ref{m2}), combined with the conservation conditions of Eq.~(\ref{cons}), we can express the time derivative of the Palatini scalar curvature as~\cite{Cao:2017jsi,Fay:2007gg}

\begin{equation}\label{m3}
\mathcal{\dot R}=-\frac{3H\left(f'\left(\mathcal{R}\right)\mathcal{R}-2f\left(\mathcal{R}\right)\right)}{f''\left(\mathcal{R}\right)\mathcal{R}-f'\left(\mathcal{R}\right)}.
\end{equation}

\noindent Using Eq.~(\ref{m3}) we can replace $\mathcal{\dot R}$ in Eq.~(\ref{m2}) to obtain~\cite{Cao:2017jsi,Sotiriou:2008rp}

\begin{equation}\label{hub}
H=\sqrt{\frac{2\kappa\left(\rho_{m}+\rho_{r}\right)+f'\left(\mathcal{R}\right)\mathcal{R}-f\left(\mathcal{R}\right)}{6f'\left(\mathcal{R}\right)\xi}},
\end{equation}

\noindent where $\xi$ is defined by

\begin{equation}\label{xi}
\xi=\left(1-\frac{3}{2}\frac{f''\left(\mathcal{R}\right)\left(f'\left(\mathcal{R}\right)\mathcal{R}-2f\left(\mathcal{R}\right)\right)}{f'\left(\mathcal{R}\right)\left(f''\left(\mathcal{R}\right)\mathcal{R}-f'\left(\mathcal{R}\right)\right)}\right)^{2}.
\end{equation}

\noindent If $\rho_{r}=0$, it is possible to use $T=-\rho_{m}=\left(f'\left(\mathcal{R}\right)\mathcal{R}-2f\left(\mathcal{R}\right)\right)/ \kappa$ to obtain the simpler expression

\begin{equation}\label{hub2}
H=\sqrt{\frac{3f\left(\mathcal{R}\right)-f'\left(\mathcal{R}\right)\mathcal{R}}{6f'\left(\mathcal{R}\right) \xi}}.
\end{equation}

In this work, we shall perform a detailed analysis of the phase space of the model defined by Eq.~(\ref{LagrangianSimple}). To facilitate this analysis we establish an autonomous system of equations defined by the pair of dimensionless variables~\cite{Fay:2007gg} 

\begin{equation}\label{variables}
y_{1}=\frac{f'\left(\mathcal{R}\right)\mathcal{R}-f\left(\mathcal{R}\right)}{6f'\left(\mathcal{R}\right)\xi H^{2}}, \qquad y_{2}=\frac{\kappa \rho_{r}}{3f'\left(\mathcal{R}\right)\xi H^{2}}.
\end{equation}

\noindent Using Eqs.~(\ref{variables}) and~(\ref{hub}) it can be shown that $\rho_{r}$ can be expressed in terms of the variable $y_{2}$ via

\begin{equation}\label{rhoR}
\rho_{r}=\frac{y_{2}}{1-y_{2}}\left(\rho_{m}+\frac{f'\left(\mathcal{R}\right)\mathcal{R}}{2\kappa}-\frac{f\left(\mathcal{R}\right)}{2\kappa}\right).  
\end{equation}

\noindent The evolution of $y_{1}$ and $y_{2}$ as a function of the cosmic scale factor $a$ are established by the differential equations 

\begin{equation}\label{diff1}
\frac{dy_{1}}{dN}=y_{1}\left(3-3y_{1}+y_{2}-3\frac{\left(f'\left(\mathcal{R}\right)\mathcal{R}-2f\left(\mathcal{R}\right)\right)f''\left(\mathcal{R}\right)\mathcal{R}}{\left(f'\left(\mathcal{R}\right)\mathcal{R}-f\left(\mathcal{R}\right)\right)\left(f''\left(\mathcal{R}\right)\mathcal{R}-f'\left(\mathcal{R}\right)\right)}\left(1-y_{1}\right)\right)
\end{equation}

\noindent and

\begin{equation}\label{diff2}
\frac{dy_{2}}{dN}=y_{2}\left(-1-3y_{1}+y_{2}+ 3\frac{\left(f'\left(\mathcal{R}\right)\mathcal{R}-2f\left(\mathcal{R}\right)\right)f''\left(\mathcal{R}\right)\mathcal{R}}{\left(f'\left(\mathcal{R}\right)\mathcal{R}-f\left(\mathcal{R}\right)\right)\left(f''\left(\mathcal{R}\right)\mathcal{R}-f'\left(\mathcal{R}\right)\right)}y_{1}\right),
\end{equation}

\noindent where $N\equiv \rm{ln}(a)$. The fixed points of this system correspond to the values $\left(y_{1},y_{2}\right)$ that satisfy

\begin{equation}
\frac{dy_{1}}{dN}=\frac{dy_{2}}{dN}=0.
\end{equation}

\noindent Note that there is a direct relationship between $\mathcal{R}$ and the variables $\left(y_{1},y_{2}\right)$ given by~\cite{Fay:2007gg}

\begin{equation}
\frac{f'\left(\mathcal{R}\right)\mathcal{R}-2f\left(\mathcal{R}\right)}{f'\left(\mathcal{R}\right)\mathcal{R}-f\left(\mathcal{R}\right)}=-\frac{1-y_{1}-y_{2}}{2y_{1}}.
\end{equation}

By calculating the eigenvalues $\left(\lambda_{1},\lambda_{2}\right)$ of the Jacobian matrix at each point $\left(y_{1},y_{2}\right)$ the stability of the fixed points can be determined~\cite{Fay:2007gg,Copeland:1997et}. The fixed point is stable when both eigenvalues are real and negative, and unstable when both are real and positive. The fixed point is a saddle point when both eigenvalues are real and of opposite sign. The nature of the fixed point for different eigenvalues $\left(\lambda_{1},\lambda_{2}\right)$ is summarized in Tab.~\ref{tab1}.

\begin{table}[H]
\begin{center}
\begin{tabular} {|c||c|}
\hline
Eigenvalues & Fixed point \\ \hline
\hline
$\lambda_{1}\neq\lambda_{2} <0$  & Stable \\ \hline
$\lambda_{1}\neq\lambda_{2} >0$ & Unstable \\ \hline
$\lambda_{1}<0<\lambda_{2}$ & Saddle \\ \hline
\end{tabular}
\end{center}
\caption{Fixed point type based on eigenvalue pairs $\left(\lambda_{1},\lambda_{2}\right)$.}  
\label{tab1}
\end{table}

The values $\left(y_{1},y_{2}\right)$ for each corresponding fixed point are then substituted into the effective equation of state $w_{eff}$ given by~\cite{Fay:2007gg}\interfootnotelinepenalty=10000 \footnote{\scriptsize As a cross-check of our methodology and computer code we verified that we are able to successfully reproduce the cosmological dynamics found in Ref.~\cite{Fay:2007gg} for two different models.}

\begin{equation}\label{weff}
w_{eff}=-y_{1}+\frac{1}{3}y_{2}+\frac{\dot f'\left(\mathcal{R}\right)}{3Hf'\left(\mathcal{R}\right)}+ \frac{\dot \xi}{3H \xi} - \frac{\dot f'\left(\mathcal{R}\right) \mathcal{R}}{18f'\left(\mathcal{R}\right) \xi H^{3}},
\end{equation}

\noindent where $\dot \xi$ is determined by taking the derivative of Eq.~(\ref{xi}) with respect to time and using Eq.~(\ref{m3}). $\dot f'\left(\mathcal{R}\right)$ is given by~\cite{Fay:2007gg}

\begin{equation}\label{m4}
\dot f'\left(\mathcal{R}\right)=-\frac{3H\left(f'\left(\mathcal{R}\right)\mathcal{R}-2f\left(\mathcal{R}\right)\right)f''\left(\mathcal{R}\right)}{f''\left(\mathcal{R}\right)\mathcal{R}-f'\left(\mathcal{R}\right)}=\mathcal{\dot R}f''\left(\mathcal{R}\right).
\end{equation}

It will also prove useful to define the deceleration parameter $q$ in terms of the effective equation of state $w_{eff}$. Since the deceleration parameter is defined in terms of the Hubble parameter via

\begin{equation}\label{m5}
q\equiv -\left(\frac{\dot H}{H^{2}} +1\right),
\end{equation}

\noindent and since~\cite{Fay:2007gg}

\begin{equation}\label{m6}
\frac{\dot H}{H^{2}}=-\frac{3}{2}\left(1+w_{eff}\right),
\end{equation}

\noindent we then find

\begin{equation}\label{m7}
q=\frac{1}{2}\left(1+3 w_{eff}\right).
\end{equation}

To further evaluate the viability criteria set out in the introduction, we must also test whether our theory contains scalar curvature singularities~\cite{Coumbe:2019fht}. A local rescaling of the metric tensor by a conformal factor $\Omega^{2}(x)$ is equivalent to the transformations~\cite{Magnano:1993bd,Higgs:1959jua,Barrow:1988xh}

\begin{equation}\label{cs1}
g_{\mu\nu} \rightarrow \tilde{g}_{\mu\nu}=f'(\mathcal{R}) g_{\mu\nu}, \qquad g^{\mu\nu} \rightarrow \tilde{g}^{\mu\nu}=\left(f'(\mathcal{R})\right)^{-1}g^{\mu\nu}.
\end{equation}

\noindent The Ricci scalar $\mathcal{R}$ defines the simplest possible curvature invariant. Thus, in the Palatini formalism, $\mathcal{R}$ raised to the power of any positive integer $m$ transforms under~(\ref{cs1}) via 

\begin{equation}\label{ricci}
\mathcal{R}^{m} \to \frac{\mathcal{R}^{m}}{\left(f'\left(\mathcal{R}\right)\right)^{m}}.
\end{equation}

The next simplest curvature invariant involves the Ricci tensor. Since our model is defined in the Palatini variation, the connection $\Gamma^{\nu}_{\mu\sigma}$ is not assumed to depend on the metric $g_{\mu\nu}$, and so the Ricci tensor

\begin{equation}
R_{\mu\nu}=\partial_{\rho}\Gamma^{\rho}_{\nu\mu} - \partial_{\nu}\Gamma^{\rho}_{\rho\mu} + \Gamma^{\rho}_{\rho\lambda}\Gamma^{\lambda}_{\nu\mu} - \Gamma^{\rho}_{\nu\lambda}\Gamma^{\lambda}_{\rho\mu}
\end{equation}

\noindent may remain invariant under the local rescaling transformation of Eq.~(\ref{cs1}). The Ricci tensor with upper indices, however, is given by $R^{\mu\nu}=g^{\mu \rho}g^{\nu \sigma}R_{\rho \sigma}$ and so it does transform under Eq.~(\ref{cs1}) according to $R^{\mu\nu} \to R^{\mu\nu}\ \left(f'(\mathcal{R})\right)^{-2}$. Therefore, second order curvature invariants involving the Ricci tensor, namely $R_{\mu\nu}R^{\mu\nu}$, to any integer power $m$, will transform under Eq.~(\ref{cs1}) according to 

\begin{equation}\label{tensor}
  \left(R_{\mu\nu}R^{\mu\nu}\right)^{m} \to \frac{\left(R_{\mu\nu}R^{\mu\nu}\right)^{m}}{\left(f'(\mathcal{R})\right)^{2m}}.
\end{equation}

\noindent It is unclear whether the Kretchmann scalar is a scalar in the Palatini formalism~\cite{Bejarano:2019zco}, and so we omit this from our analysis.  

\end{section}


\begin{section}{Results}\label{results}

  We find that the model defined by the exponent of Eq.~(\ref{LagrangianSimple}) contains three fixed points $P_{1}$, $P_{2}$ and $P_{3}$. The eigenvalues and stability of these fixed points, defined by the roots $\left(y_{1},y_{2}\right)$ of Eqs.~(\ref{diff1}) and~(\ref{diff2}), are displayed in Tab.~\ref{tab2} in the low and high-curvature limits. Figure~\ref{OP} displays how the eigenvalues $\left(\lambda_{1},\lambda_{2}\right)$ vary as a function of $\mathcal{R}_{*}$ for $P_{1}$ (left), $P_{2}$ (middle), and $P_{3}$ (right).
  
\begin{table}[H]
\begin{center}
\begin{tabular} {|c|c|l|l|}
\hline
{Fixed point} & $\left(y_{1},y_{2}\right)$ & $\left(\lambda_{1},\lambda_{2}\right)$ $\left(\mathcal{R}_{*}\to 0\right)$ & $\left(\lambda_{1},\lambda_{2}\right)$ $\left(\mathcal{R}_{*}\to 1\right)$ \\ \hline
\hline
$P_{1}$ & $\left(1,0\right)$ & $\left(6,5\right)$ Unstable & $\left(-4,-3\right)$ Stable \\ \hline
$P_{2}$ & $\left(0,1\right)$ & $\left(1,-5\right)$ Saddle & $\left(1,4\right)$ Unstable \\ \hline
$P_{3}$ & $\left(0,0\right)$ & $\left(-1,-6\right)$ Stable & $\left(-1,3\right)$ Saddle \\ \hline
\end{tabular}
\end{center}
\caption{The dimensionless variables $\left(y_{1},y_{2}\right)$, eigenvalues $\left(\lambda_{1},\lambda_{2}\right)$ in the low $\left(\mathcal{R}_{*}\to 0\right)$ and high-curvature $\left(\mathcal{R}_{*}\to 1\right)$ limits, and stability of the three fixed points $P_{1}$, $P_{2}$ and $P_{3}$.}
\label{tab2}
\end{table}   

\begin{figure}[h!]
\centering
\scalebox{.6}{\includegraphics{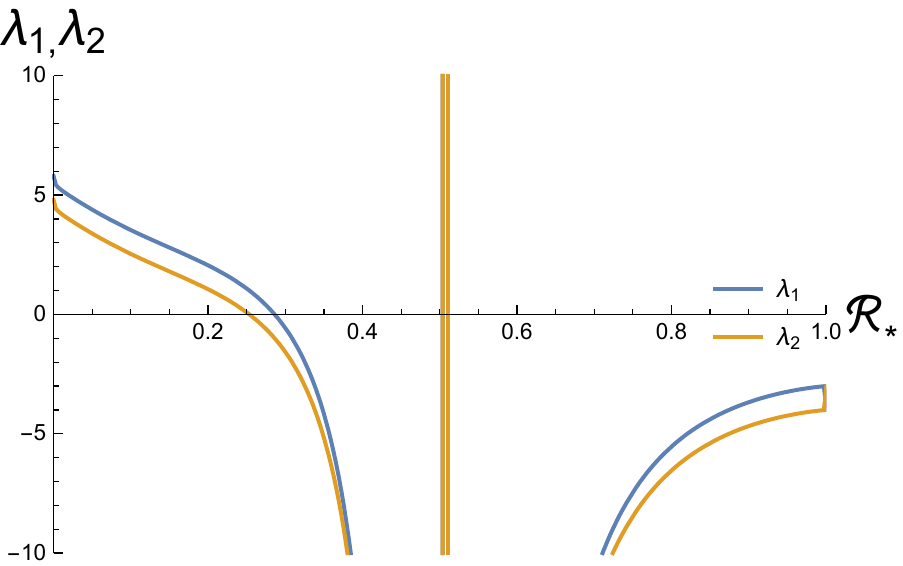}}
\scalebox{.6}{\includegraphics{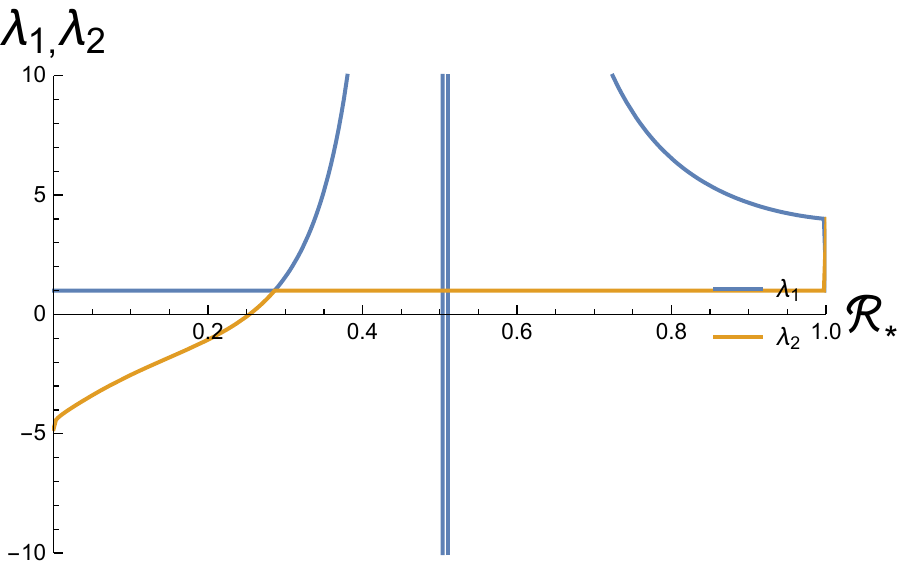}}
\scalebox{.6}{\includegraphics{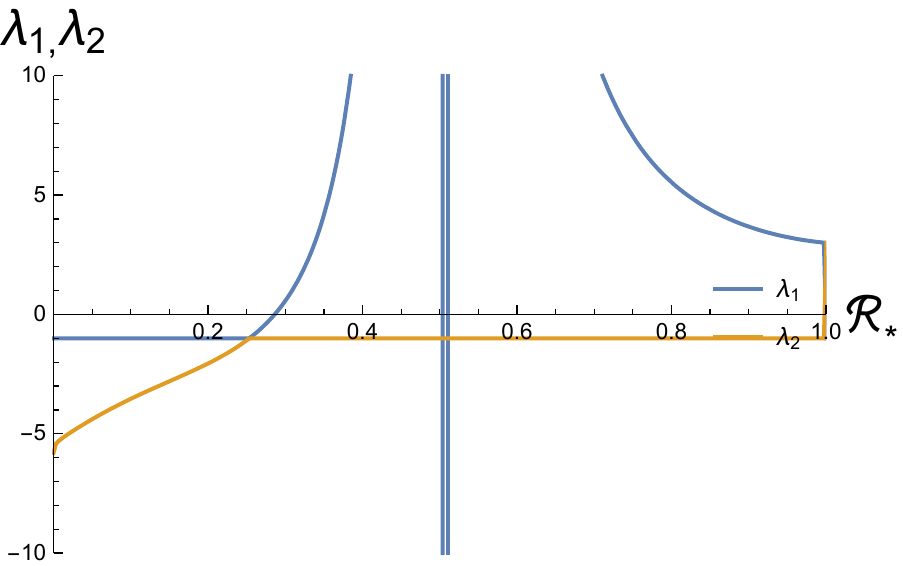}}
\caption{The eigenvalues $\left(\lambda_{1},\lambda_{2}\right)$ as a function of $\mathcal{R}_{*}$ for fixed points $P_{1}$ (left), $P_{2}$ (middle), and $P_{3}$ (right).}
\label{OP}
\end{figure}

Figure~\ref{OP} illustrates a potential advantage of AWIG. Unlike most other $f(R)$ theories of gravity, the variable power in the Lagrangian density of AWIG makes it possible for the eigenvalues and hence stability of each fixed point to vary with curvature scale, and hence to potentially vary with cosmological time. So, for example, the stability of the fixed point $P_{1}$ can change from being stable in the high-curvature limit to being unstable at lower curvatures, as can be seen in Fig.~\ref{OP} (left). This feature allows a richer set of possible cosmological dynamics.     

Inserting the obtained coordinate pairs $\left(y_{1},y_{2}\right)$ into Eq.~(\ref{weff}) yields the effective equation of state $w_{eff}$ as a function of the Palatini scalar curvature. The results are displayed in Fig.~\ref{figweff} for the fixed points $P_{1}$ and $P_{3}$. Since $y_{2}=1$ for the fixed point $P_{2}$ we can see from Eq.~(\ref{rhoR}) that $\rho_{r}$ is undefined, and therefore via 
Eq.~(\ref{hub}) $H$ must also be undefined. Consequently, $w_{eff}$ for $P_{2}$ is undefined, as is evident from Eq.~(\ref{weff}). However, we know that as $n\left(\mathcal{R}_{*}\right) \to 2$ AWIG is equivalent to general relativity plus a cosmological constant~\cite{Edery:2019txq}. Thus, $w_{eff}$ for $P_{2}$ can be determined in the high-curvature limit by an equivalent analysis of the model $f\left(\mathcal{R}\right)=\mathcal{R}-\Lambda$, where $\Lambda$ is the cosmological constant. We have repeated the methodology outlined in section~\ref{method} for the model $f\left(\mathcal{R}\right)=\mathcal{R}-\Lambda$ finding eigenvalues $\left(\lambda_{1},\lambda_{2}\right)=\left(1,4\right)$, which agrees with our result presented in Fig.~\ref{OP} (middle) in the high-curvature limit, and an effective equation of state $w_{eff}=1/3$. Identical results are also found in Ref.~\cite{Fay:2007gg}. Therefore, $P_{2}$ corresponds to a radiation-like phase in the high-curvature limit.

\begin{figure}[H]
  \centering
\scalebox{.45}{\includegraphics{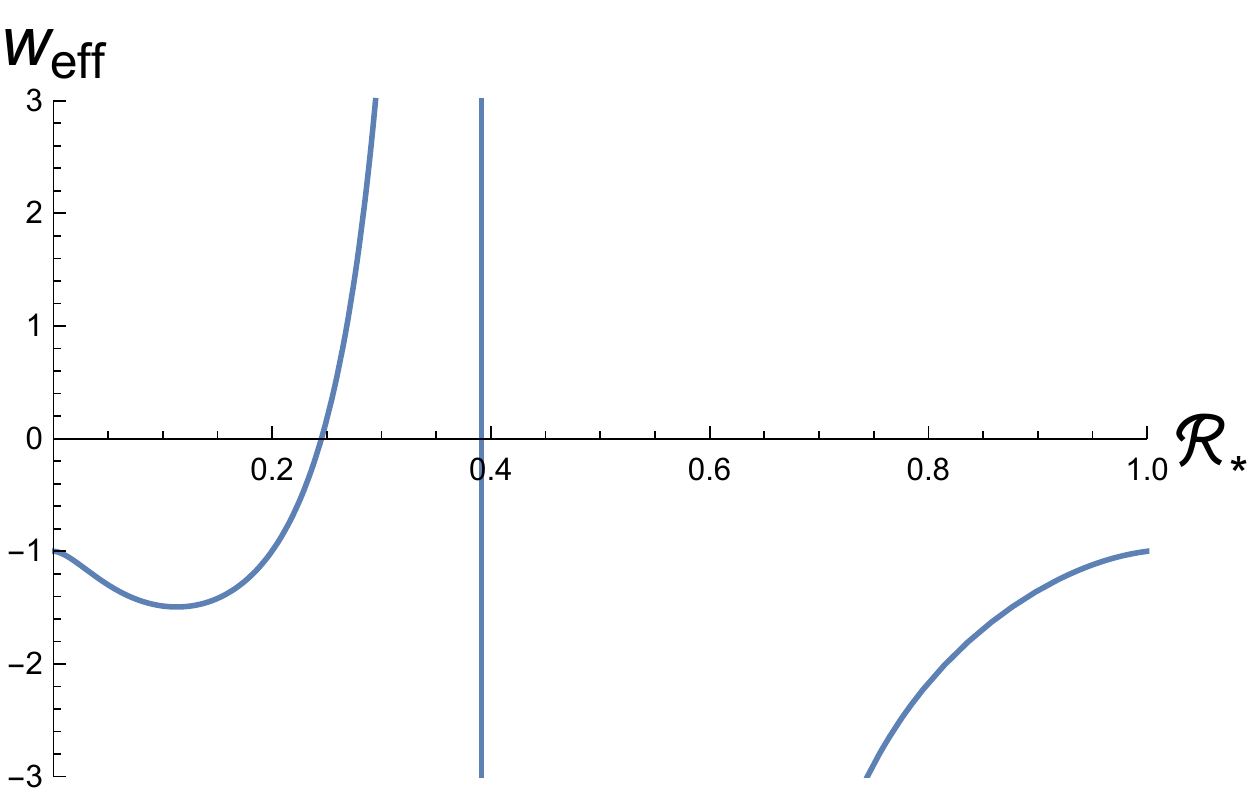}}
\scalebox{.45}{\includegraphics{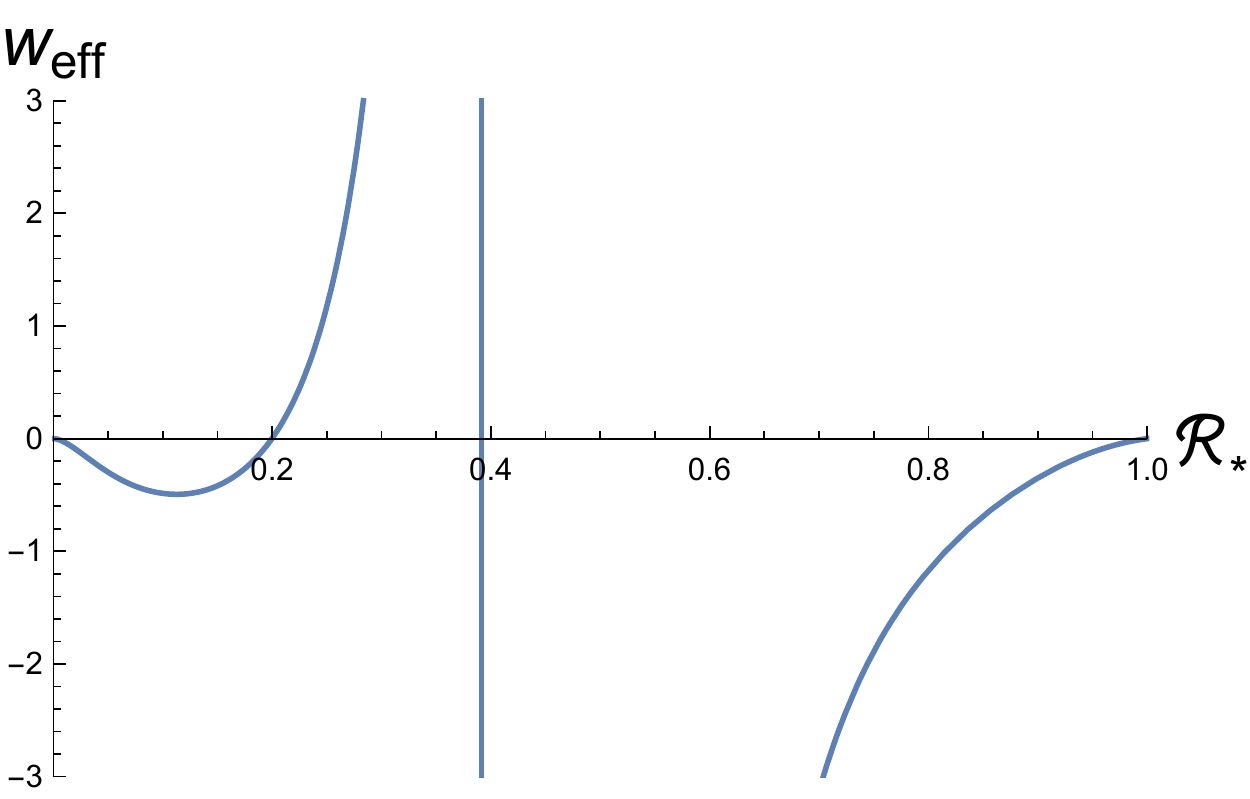}}
\caption{The effective equation of state parameter $w_{eff}$ as a function of $\mathcal{R}_{*}$ for the fixed point $P_{1}$ (left) and $P_{3}$ (right).}
\label{figweff}
\end{figure}

The effective equation of state parameter $w_{eff}$ for the fixed points $P_{1}$, $P_{2}$ and $P_{3}$ in the low and high-curvature limits are summarised in Tab.\ref{tab3}. Thus, we identify $P_{1}$ as a de Sitter-like phase, $P_{2}$ as a radiation-like phase, and $P_{3}$ as a matter-like phase. Note that the unknown value of $w_{eff}$ for $P_{2}$ in the limit $\mathcal{R}_{*}\to 0$ is denoted by $-$. Figures~\ref{OP} and~\ref{figweff} suggest that if the matter-dominated phase $P_{3}$ is to transition back to the de Sitter-like phase $P_{1}$, to account for the currently observed late period of cosmic acceleration, then this transition must occur at a curvature scale $\mathcal{R}_{*} \gtrsim 0.28$. This is because if $\mathcal{R}_{*} \lesssim 0.28$ then it is not possible to exit the stable matter-like phase.

\begin{table}[H]
\begin{center}
\begin{tabular} {|c|c|c|c|}
\hline
{Fixed point} & $w_{eff}\left(\mathcal{R_{*}} \to 0 \right)$ & $w_{eff}\left(\mathcal{R_{*}} \to 1 \right)$ & {Phase} \\ \hline
\hline
$P_{1}$ & -1 & -1 & {De Sitter} \\ \hline
$P_{2}$ & - & 1/3 & {Radiation} \\ \hline
$P_{3}$ & 0 & 0 & {Matter} \\ \hline
\end{tabular}
\end{center}
\caption{The effective equation of state in the low-curvature limit $w_{eff}\left(\mathcal{R}_{*} \to 0 \right)$, high-curvature limit $w_{eff}\left(\mathcal{R}_{*} \to 1 \right)$ and the phase type for the fixed points $P_{1}$, $P_{2}$ and $P_{3}$.}
\label{tab3}
\end{table}  

To further analyse the cosmological evolution of our model we use Eq.~(\ref{m7}) to investigate how the deceleration parameter $q$ varies as a function of $\mathcal{R}_{*}$ for the de Sitter-like phase. The results are shown in Fig.~\ref{figweff3}. If $q>0$ then the universe is expanding but decelerating. If $q<0$ then the universe is expanding but accelerating~\cite{Bolotin:2015dja}. Figure~\ref{figweff3}, therefore, indicates that the de Sitter-like phase undergoes two periods of accelerated expansion, one in the high-curvature regime $0.4 \lesssim \mathcal{R} < 1$ and one in the low-curvature regime $0 \leq \mathcal{R} \lesssim 0.23$, mediated by a period of decelerated expansion for $0.23 \lesssim \mathcal{R} \lesssim 0.4$ (see Fig.~\ref{figweff3}). Assuming curvature on cosmological scales decreases with cosmological time, this implies an early and late period of accelerated cosmic expansion, with an intermediate period of decelerated expansion. In this sense, the dynamics appear consistent with cosmological observations, depending on the exact scale set by $\mathcal{R}_{0}$.

\begin{figure}[H]
  \centering
\scalebox{0.7}{\includegraphics{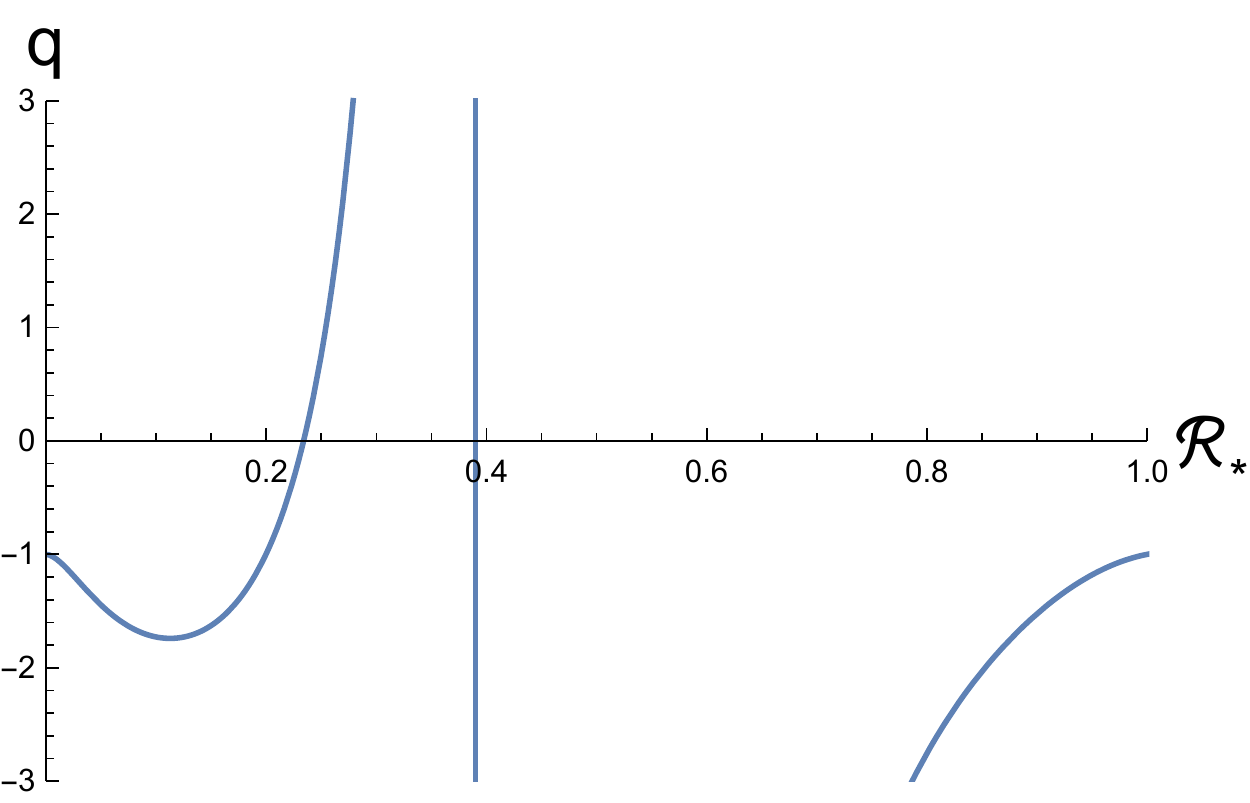}}
\caption{The deceleration parameter $q$ as a function of scalar curvature $\mathcal{R}_{*}$ for the fixed point $P_{1}$.}
\label{figweff3}
\end{figure}

We now analyse the phase space of this model, with the results shown in Fig.~\ref{SP}. The phase space of AWIG is 3-dimensional, with each point in the phase space uniquely specified by the set of coordinates $\left(y_{1},y_{2},\mathcal{R}_{*}\right)$. Figure~\ref{SP} shows the $\left(y_{1},y_{2}\right)$ plane for three different values of constant curvature. One possible route the system may take through the 3-dimensional phase space is depicted in the three plots of Fig.~\ref{SP}, where the system evolves through the closed sequence of fixed points $P_{1}\to P_{2}\to P_{3}\to P_{1}$ with decreasing curvature scale $\mathcal{R}_{*}$. Thus, the model presented is consistent with the sequence of an early period of accelerated expansion, intermediate radiation and matter-dominated eras of decelerated expansion, followed by the return to a period of accelerated expansion at late times.    

\begin{figure}[H]
  \centering
\scalebox{.45}{\includegraphics{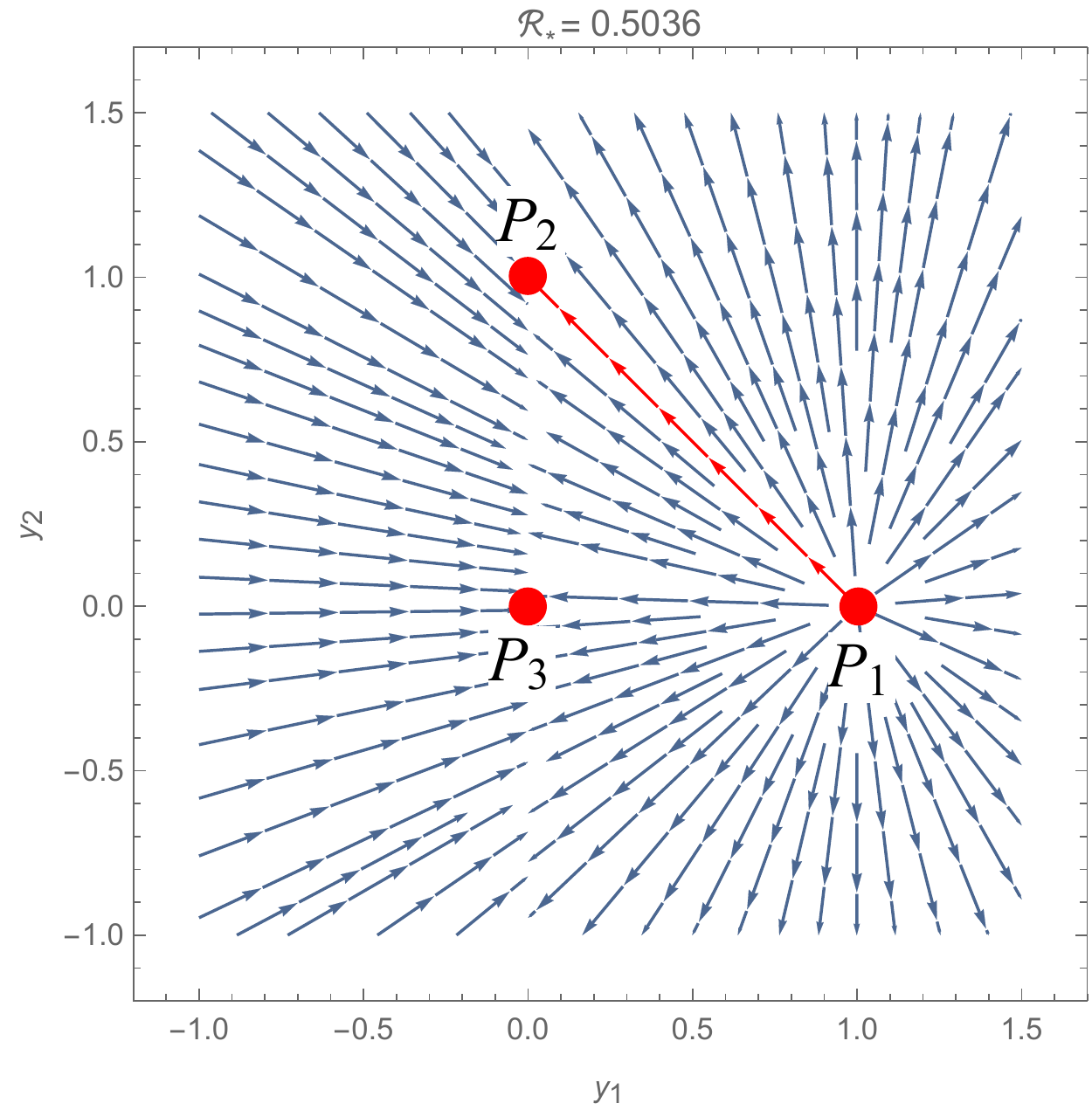}}
\scalebox{.45}{\includegraphics{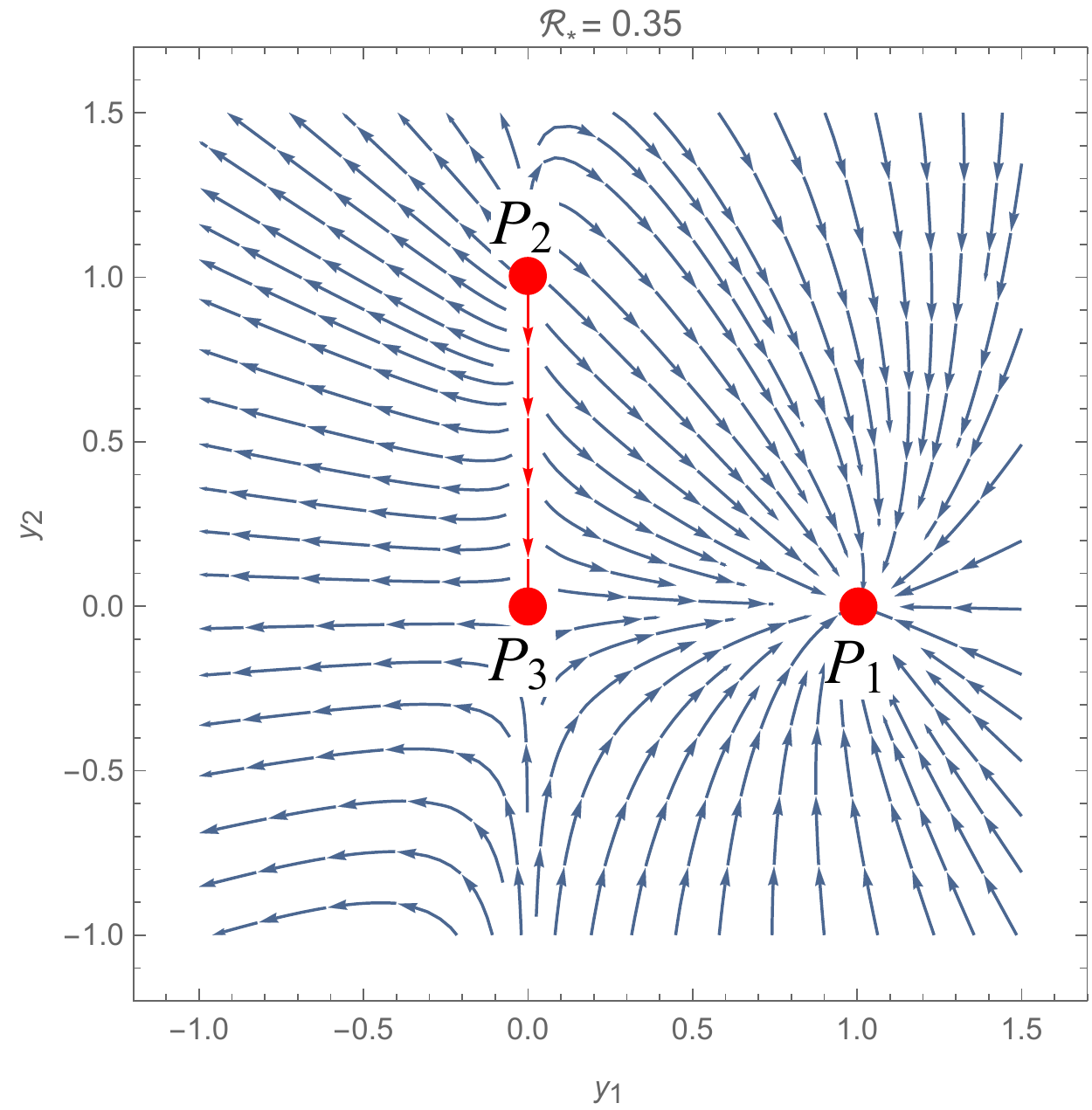}}
\scalebox{.45}{\includegraphics{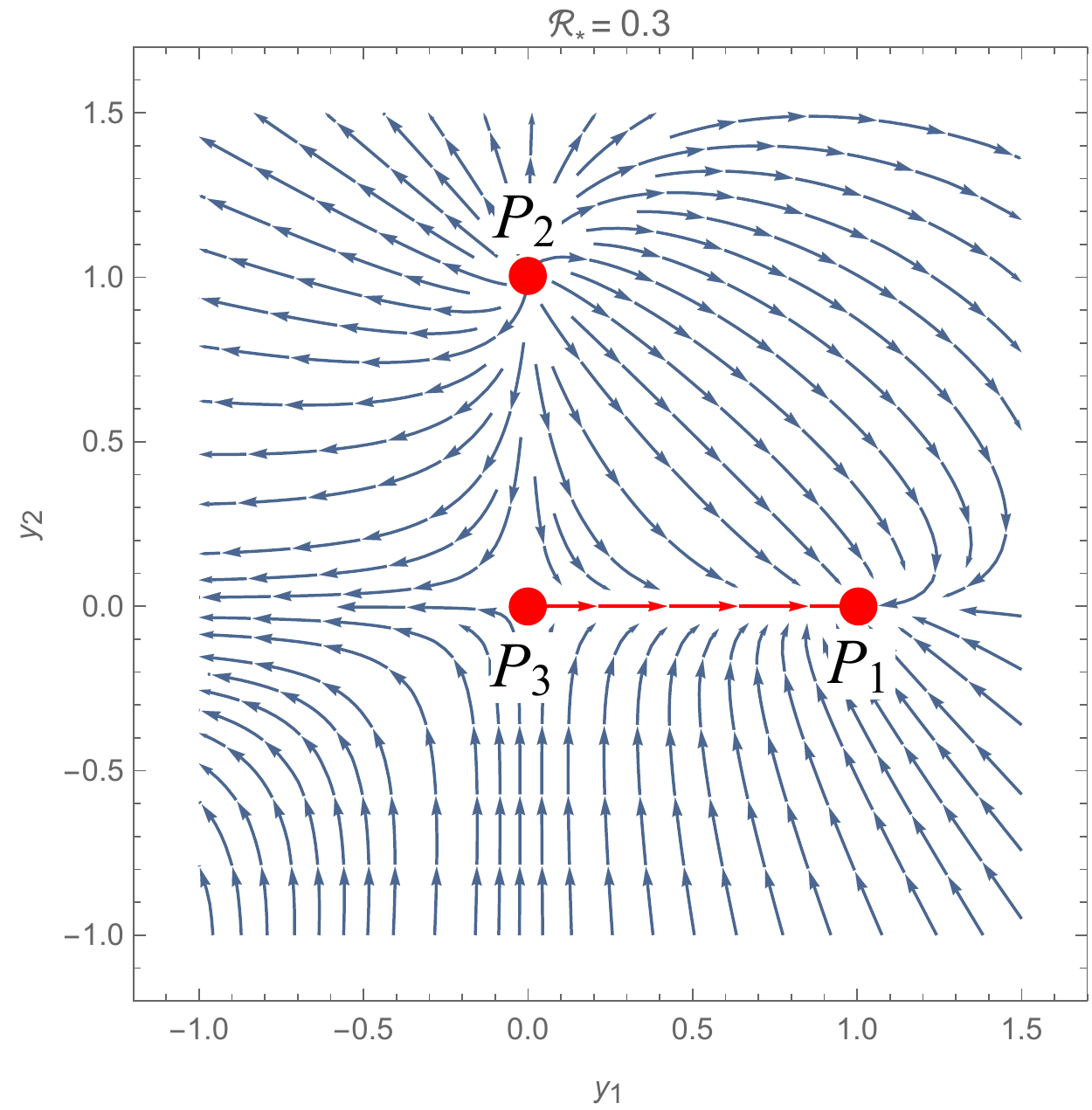}}
\caption{Slices of constant curvature through the 3-dimensional phase space of AWIG at $\mathcal{R}_{*}=0.5036$ (left), $\mathcal{R}_{*}=0.35$ (middle) and $\mathcal{R}_{*}=0.3$ (right). The red trajectory shows one possible way the system may evolve through the sequence of fixed points $P_{1}\to P_{2}\to P_{3}\to P_{1}$.}
\label{SP}
\end{figure}

We now present results for various powers of the Ricci scalar curvature under the local rescaling of Eq.~(\ref{cs1}). Using Eqs.~(\ref{fp1}) and~(\ref{ricci}) we find 

\begin{equation}
\mathcal{R}^{m} \to \frac{\mathcal{R}^{m}}{\left(f'\left(\mathcal{R}\right)\right)^{m}}\underset{\mathcal{R}_{*} \to 1}{=}\frac{1}{2^{m}}.
\end{equation}

\noindent The first three powers of the Ricci scalar curvature ($m=1,2,3$) are shown in Fig.~\ref{curvinv}. As can be seen from Fig.~\ref{curvinv} each curvature invariant is divergence-free and approaches a constant in the limit $\mathcal{R}_{*} \to 1$. Similar results have been shown in Refs.~\cite{Bambi:2015zch,Barragan:2009sq}. Likewise, Eqs.~(\ref{fp1}) and~(\ref{tensor}) can be used to show that the curvature invariant $\left(R_{\mu\nu}R^{\mu\nu}\right)^{m}$ formed from the Ricci tensor asymptotically approaches $1/2^{2m}$ as $\mathcal{R}_{*} \to 1$. Therefore, the model presented contains no curvature singularities in $\mathcal{R}$ or $R_{\mu\nu}R^{\mu\nu}$, at any order $m$.   

\begin{figure}[H]
\centering
\scalebox{1.0}{\includegraphics{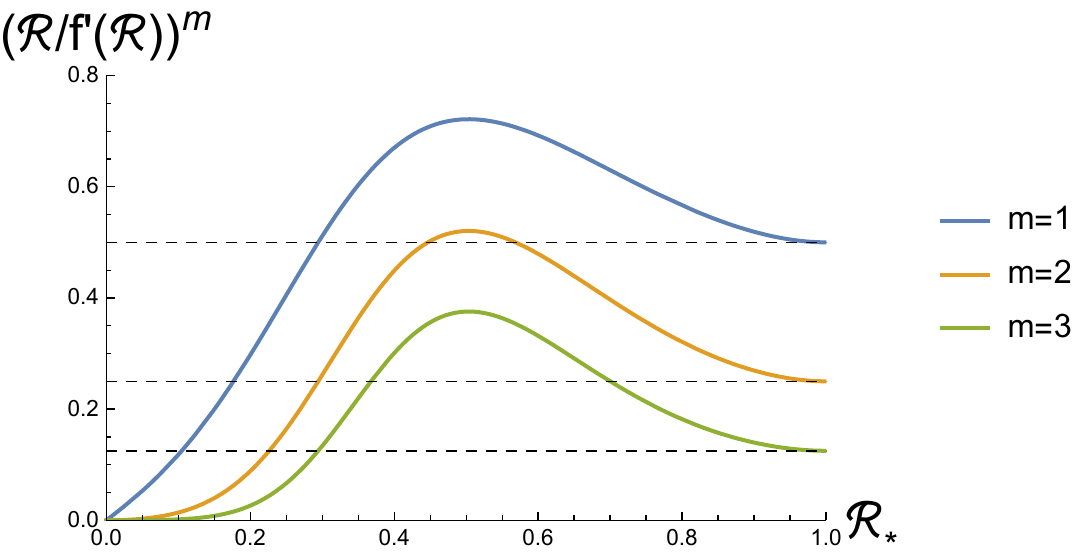}}
\caption{The first three powers ($m=1,2,3$) of the transformed Palatini scalar curvature as a function of $\mathcal{R}_{*}$.}
\label{curvinv}
\end{figure}

\end{section}

\begin{section}{Discussion}\label{discussion}

In this work we have shown that one of the simplest possible implementations of asymptotically Weyl-invariant gravity (AWIG) may be viable, as measured against criteria (\romannumeral 1)$-$(\romannumeral 6) set out in the introduction. 

However, the model's viability cannot yet be definitively established for several reasons. Firstly, AWIG is by construction superficially renormalizable, but establishing its renormalizability via explicit calculation remains an open problem. Secondly, the analysis performed in this work has raised some unanswered questions. For example, the transition from the matter-dominated phase to the late phase of cosmic expansion must occur at a curvature scale $\mathcal{R} \gtrsim 0.28 \mathcal{R}_{0}$. It is unknown whether this is consistent with cosmological observations since the dimensionful scale $\mathcal{R}_{0}$ is presently unknown. Furthermore, the effective equation of state parameter $w_{eff}$ for the fixed point $P_{3}$ is negative for $0 < \mathcal{R}_{*} \lesssim 0.2$, the meaning of which is unclear. Finally, one of the three fixed points $P_{2}$ has an undefined effective equation of state for $0 \leq \mathcal{R}_{*} < 1$, however, we can determine $w_{eff}$ for $\mathcal{R}_{*}\to 1$. 


Nevertheless, the model presented contains several encouraging features, such as the apparent absence of curvature singularities and three fixed points with effective equation of states corresponding to de Sitter, radiation and matter-like phases. The model also contains the correct sequence of early and late periods of accelerated cosmic expansion, with an intermediate period of decelerated expansion, something that has proven difficult to achieve in other attempted modifications of general relativity~\cite{Fay:2007gg}. Moreover, the early accelerating phase emerges from AWIG without adding a scalar field. This is because AWIG asymptotically approaches the Palatini formulation of pure $\mathcal{R}^{2}$ gravity in the high curvature limit, which is equivalent to general relativity plus a non-zero cosmological constant and no massless scalar field~\cite{Edery:2019txq}. Another positive feature of AWIG is that the variable power in the Lagrangian density seems to permit a richer set of possible cosmological dynamics, as can be seen from the variable eigenvalues in Fig.~(\ref{OP}).

The dimensionless exponent $n\left(\mathcal{R}_{*}\right)$ explored in this work is among the simplest possible choices, but it is far from the only consistent choice. Going forward, we aim to narrow down, or perhaps even uniquely determine, the functional form of $n\left(\mathcal{R}_{*}\right)$ and the value of $\mathcal{R}_{0}$ to more robustly test the viability of asymptotically Weyl-invariant gravity.

\end{section}

\begin{section}{Acknowledgements}

I wish to thank Roberto Percacci and Subir Sarkar for their comments on the manuscript, and the anonymous referee for invaluable corrections.    

\end{section}
  
  
\bibliographystyle{unsrt}
\bibliography{Master}

\end{document}